\documentclass[a4paper]{PoS}
\usepackage{wrapfig}
\usepackage{gensymb}

\title{The ILD/CALICE Silicon-Tungsten Electromagnetic Calorimeter: status and potential}

\ShortTitle{The ILD/CALICE Silicon-Tungsten Electromagnetic Calorimeter: status and potential}

\author{\speaker{Kostiantyn Shpak}%
         \\
          on behalf of the CALICE/ILD SiW ECAL group,\\
        LLR - Ecole polytechnique, Centre National de la Recherche Scientifique (FR)\\
        E-mail: \email{kostiantyn.shpak@cern.ch}}

\abstract{The Particle Flow Algorithms adopted for future $e^{+}e^{-}$ colliders detectors and phase-II CMS upgrade require very high granularity calorimeters to deconvolve the individual contributions of particles in jets. This is especially true for electromagnetic calorimeters (ECAL). For a realistic large detector many technological requirements have to be fulfilled: modularity for industrialization; compact integration of an embedded very front-end electronics featuring large dynamics, low-power and self-triggering; mechanical structure and cooling systems with minimal dead zones. The technological prototype of the silicon-tungsten (SiW) ECAL presented here should achieve all this. 10 layers are produced and tested in beam, while design and optimization studies are ongoing on a variety of simulated key processes to test the performance of the hardware and the algorithms.}

\FullConference{38th International Conference on High Energy Physics\\
                  3-10 August 2016\\
                 Chicago, USA}

\begin{document}

\begin{wrapfigure}{r}{0.31\textwidth}
\vspace{-6.0em}
  \includegraphics[width=.28\textwidth]{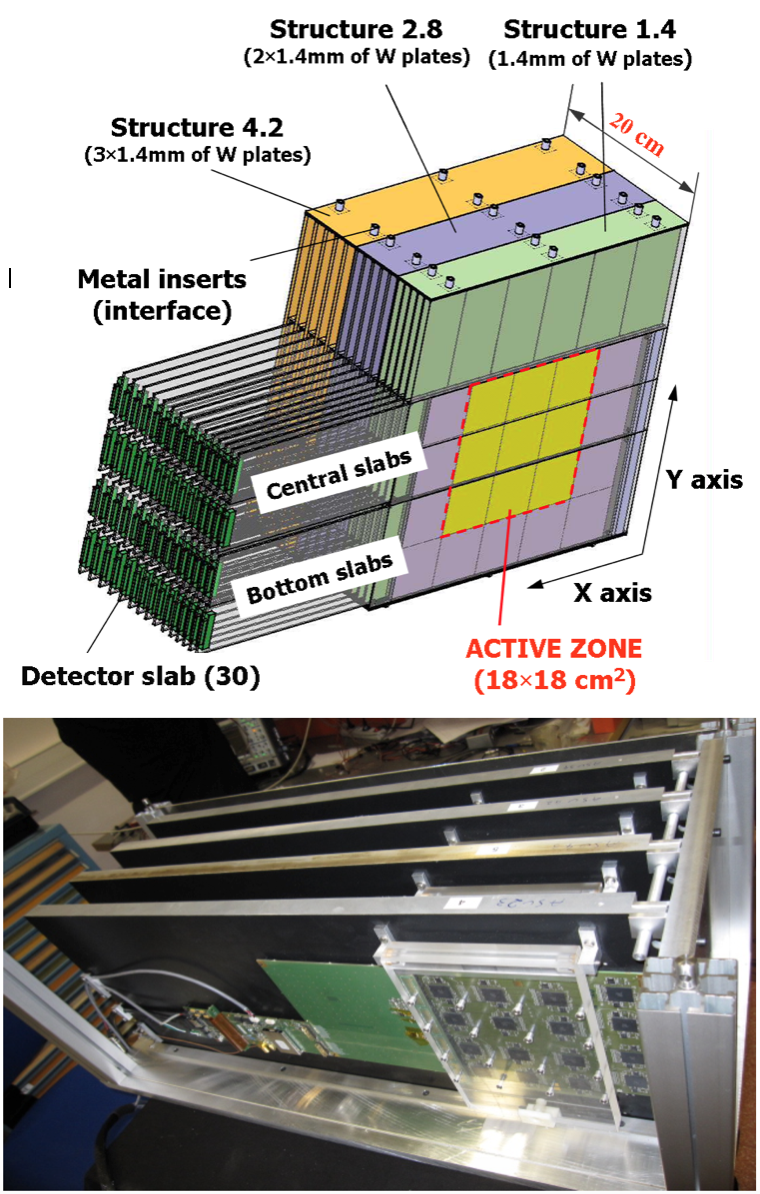}
  \vspace{-1.0em}
 \caption{Schematic 3D view of CALICE ECAL physics prototype (top) and technological prototype setup tested in November'15 at CERN (bottom).} 
 \label {calice}
\vspace{-2.0em}
\end{wrapfigure}

\section{Introduction}

CALICE SiW ECAL group is working on highly-granular electromagnetic calorimeter suitable for Particle Flow Algorithms, for example at International Large Detector (ILD) at ILC. This technology can also be used in CEPC, FCC, CMS HGCAL (approved proposal) and ATLAS HGTD. 
Depending on ILD model, ECAL has 22 or 30 layers with the silicon pixels of $5 \times 5 mm^{2}$ and the tungsten absorber (about $24$ radiation and $0.8$ interaction lengths). With PFA one should achieve $3-4\%$ jet energy resolution for $40-250GeV$ jets. This is sufficient to separate $W$ and $Z$ decaying hadronically.

\section{SiW ECAL prototypes}

CALICE ECAL physics prototype (2006-2011) was made to prove physics performance of highly granular calorimetry. Very front end (VFE) electronics was not embedded into the active layer with $1 \times 1 cm^{2}$ Si cells.

Starting from 2011 we are developing a new technological prototype. R\&D studies are concentrated on compact detector solution suitable for mass production. Four $9 \times 9 cm^{2}$ sensors with $5.5 \times 5.5 mm^{2}$ cells are glued on recently produced FEV11 printed circuit boards (PCB), equipped with 16 dedicated ASIC chip SKIROC2 with 64 channels (1024 channels per PCB) with $1fQ-10pQ$ dynamic range. Technological prototype can be operated in power pulsing mode to emulate future ILC behavior: duty cycle of ILC bunch trains is $1ms/200ms=0.5\%$, VFE electronics is switched ON $1ms$ before bunches for stabilization and OFF in ILC idle time to reduce power dissipation.

\begin{figure}
\centering
\vspace{-0.8em}
	\includegraphics[width=.47\textwidth]{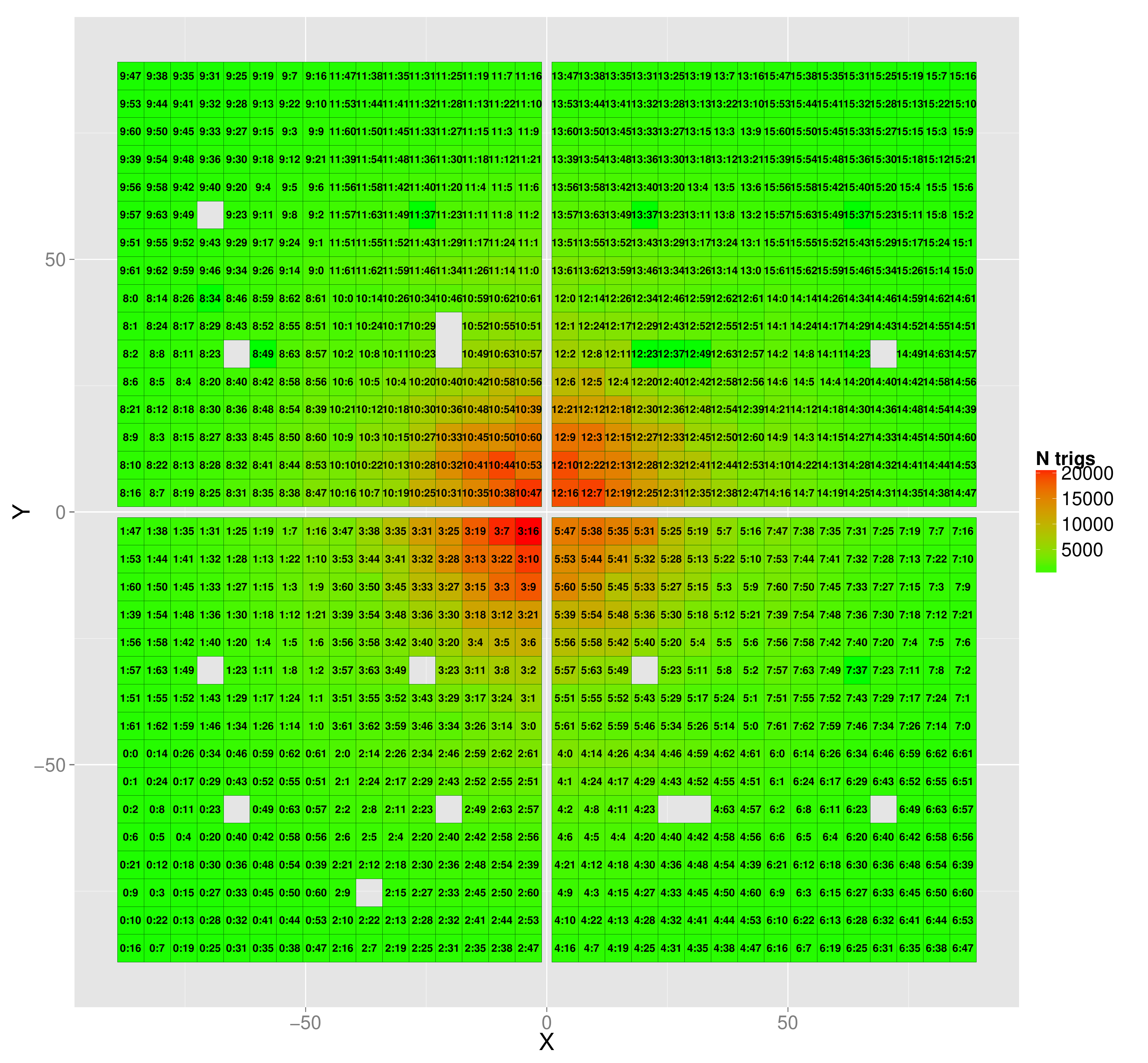}
	\includegraphics[width=.47\textwidth]{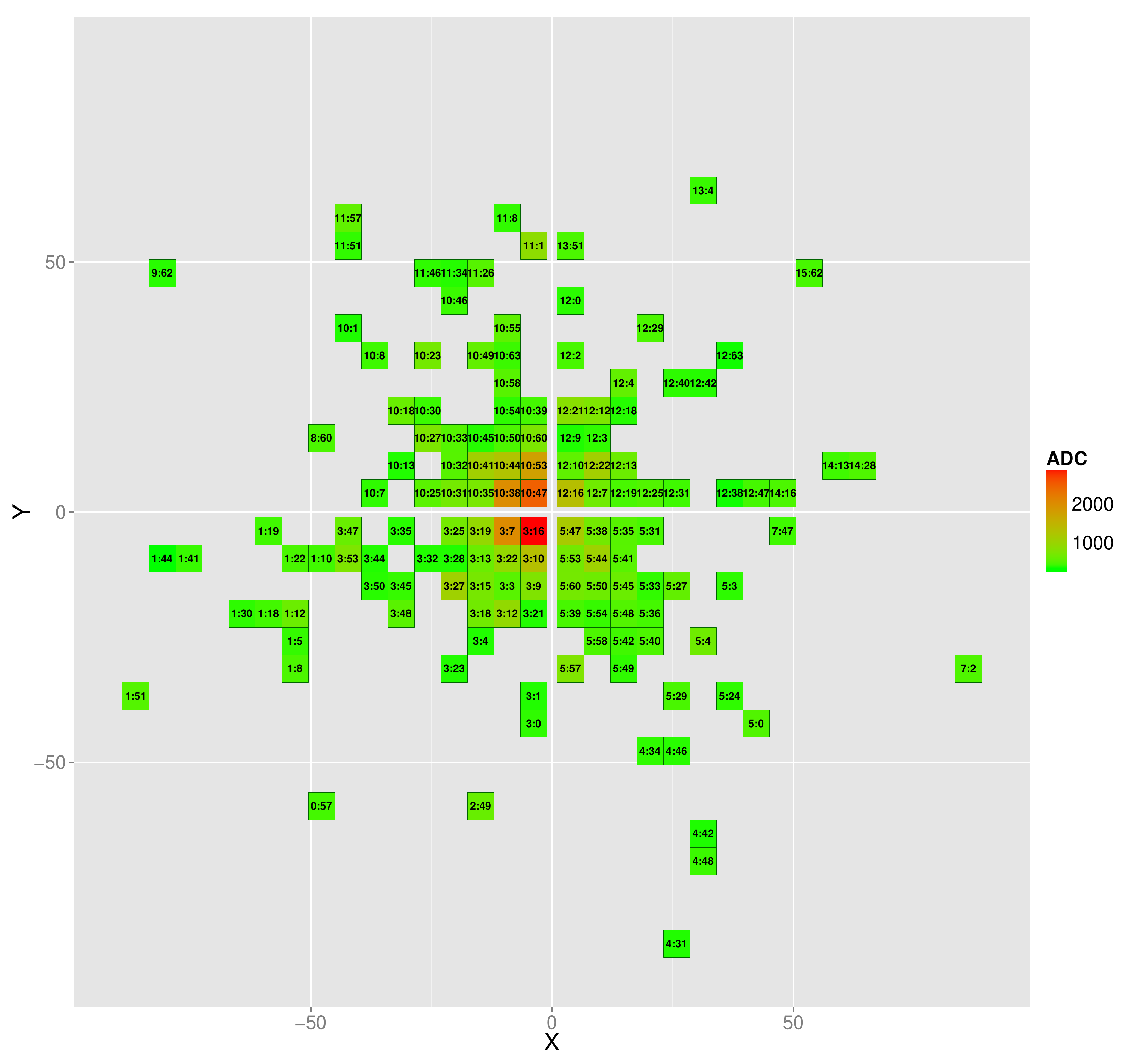}
	\vspace{-1.0em}
	\caption{Beam spot (left), example of $150GeV$ $e^{+}$ shower in one layer after $8.4X_0$ of $W$(right).}
	\label {beamspot}
	\vspace{-1.5em}
\end{figure}

\section{Technological prototype performance}

Setup equipped with four FEV11 boards was tested at CERN SPS in November'15. One layer was suffered from high noises and was switched off. Setup was showing stable behavior during all period of tests. Prototype was tested with $15$ to $150GeV$ $e^{+}, \pi^{+}, \mu^{+}$ beams with different amount of absorber in front of the setup and between the layers, at $0\degree$, $\sim 45\degree$ and $90\degree$ angles. Clear beam spot in every layer was always seen (Fig. \ref{beamspot}). $2.2\%$ of channels were masked (mostly channel 37 in every chip).

\begin{figure}[t]
\centering
\vspace{-1.5em}
	\includegraphics[width=.39\textwidth]{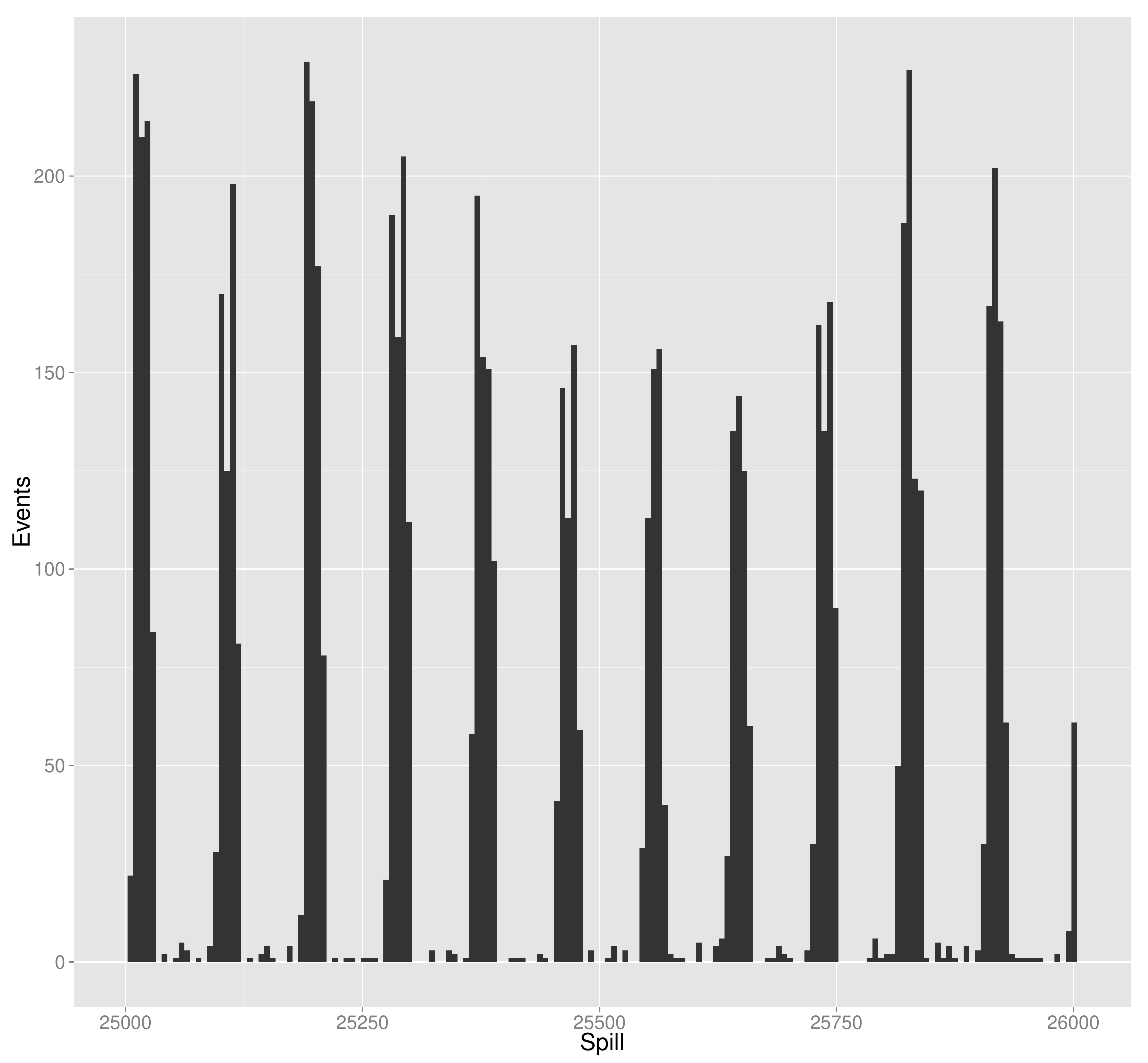}
	\includegraphics[width=.54\textwidth]{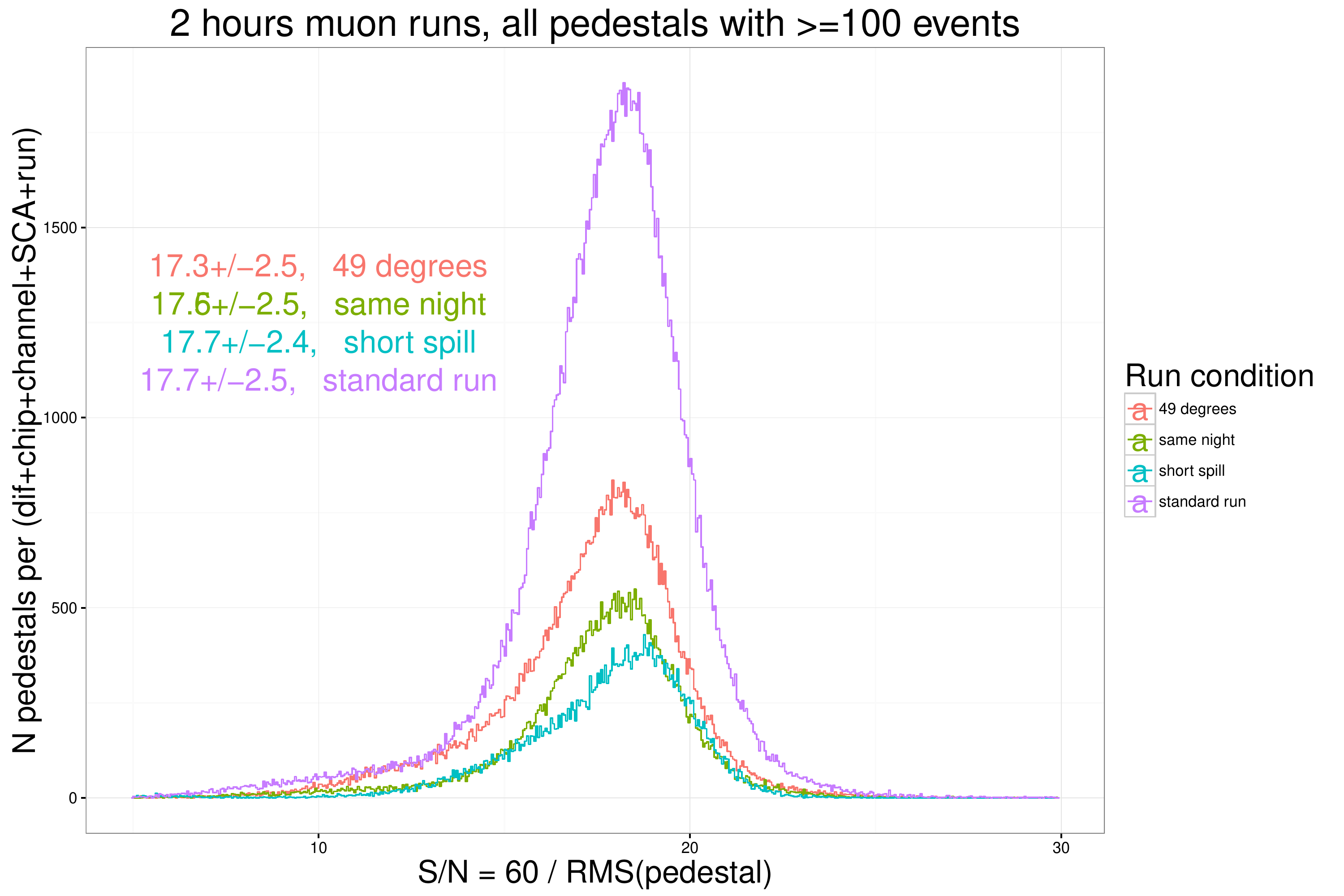}
	\vspace{-1.0em}
	\caption{Measured timing SPS spill structure (left), $\mu^{+}$ signal over pedestal RMS noise ratio (right).}
	\label {signalnoise}

	\includegraphics[width=.49\textwidth]{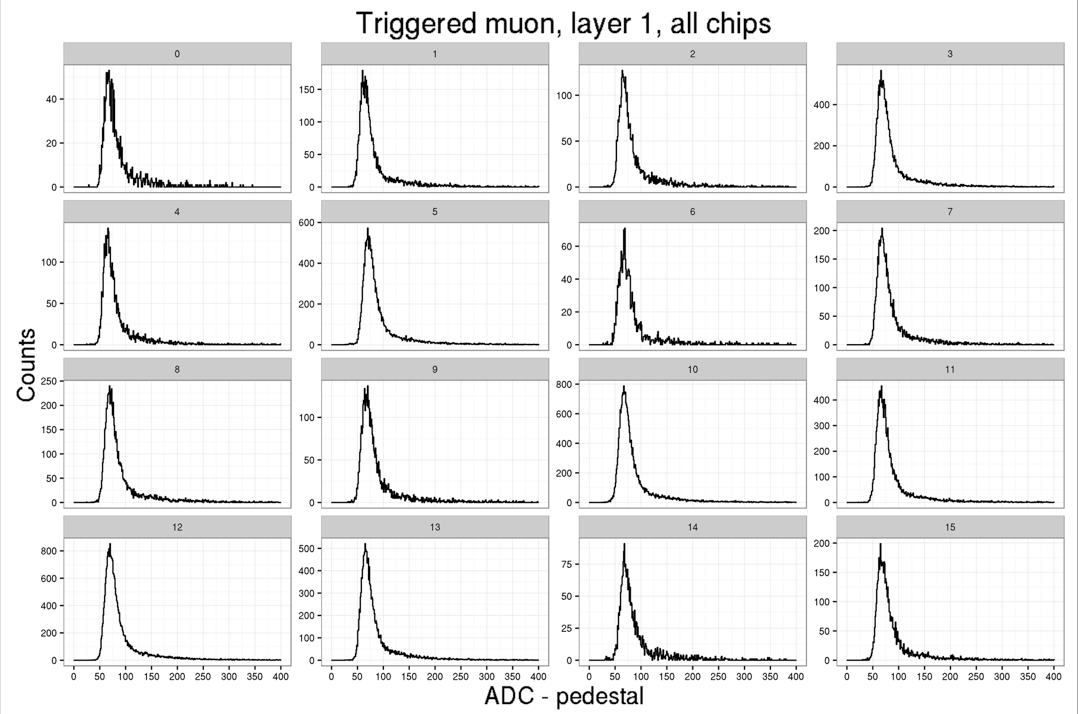}
	\includegraphics[width=.49\textwidth]{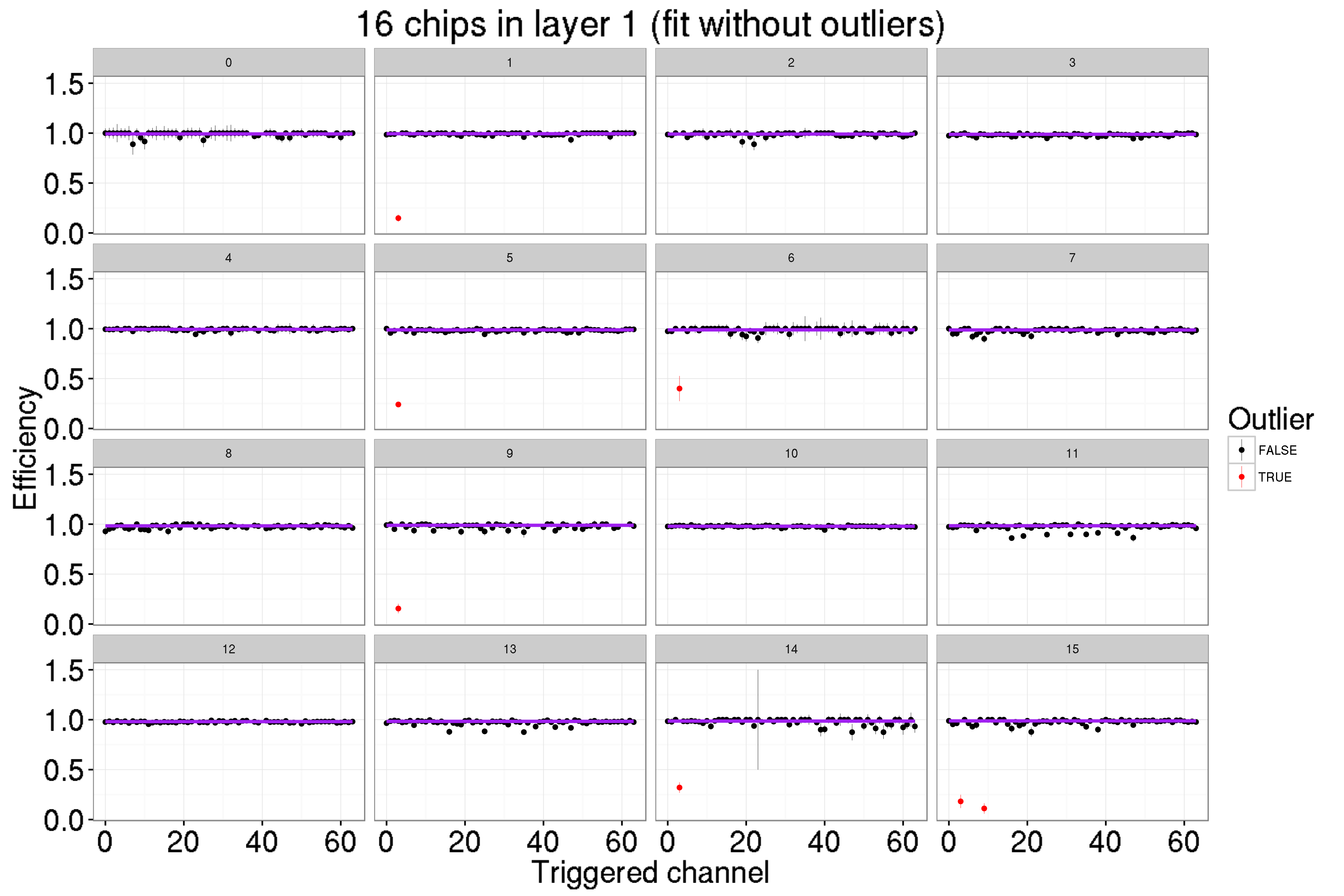}
	\vspace{-1.0em}
	\caption{Muon signals per chip (left) and efficiency per chip, channel (right) for the 1st layer (others are similar).}
	\label {muon}
	\vspace{-1.0em}
\end{figure}

\begin{wrapfigure}{r}{0.45\textwidth}
\vspace{-4.0em}
	\includegraphics[width=.45\textwidth]{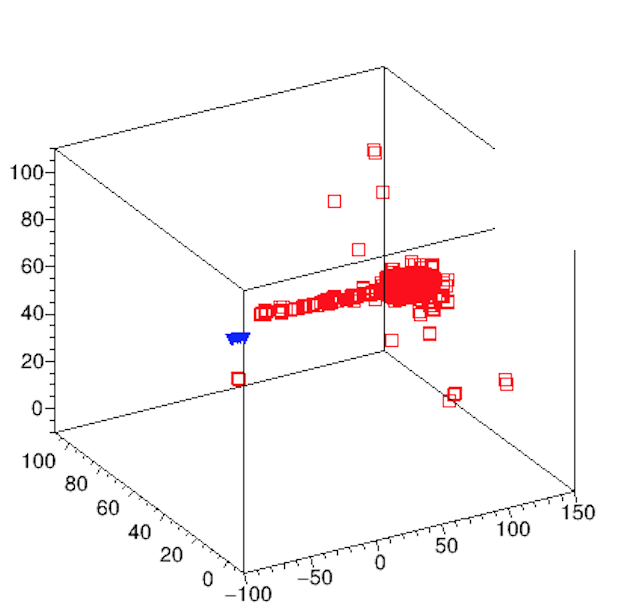}
	\vspace{-2.0em}
	\caption{Example of combined $\pi^{+}$ event (SiW ECAL hits in blue, SDHCAL in red).}
	\label {combine}
	\vspace{-1.0em}
\end{wrapfigure}
With the muon beam at normal incidence a good signal over noise ratio $S/N=17...18$ was achieved for various detector settings (Fig. \ref{signalnoise}). Setup has a very good uniformity of the muon responses across all channels ($\simeq 5\%$), its variation is dominated by VFE electronics. An average muon efficiency is about $98-99\%$ (Fig. \ref{muon}).

Combined SiW ECAL (10 layers) and full SDHCAL prototype setup was tested in June'16 at CERN SPS. Synchronization and common DAQ were tested (Fig. \ref{combine}). Unfortunately high noises were observed in ECAL during the tests.

\begin{wrapfigure}{r}{0.45\textwidth}
	\vspace{-6.5em}
	\includegraphics[width=.45\textwidth]{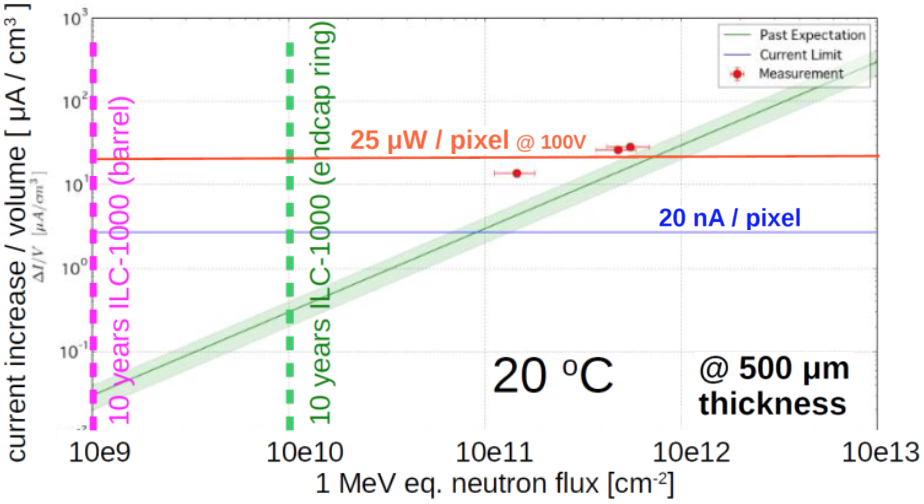}
	\vspace{-2.0em}
	\caption{Increase of dark current  is acceptable for 10 years of ILC operation at $\sqrt{s}=1 TeV$.}
	\label {darkcurrent}
	\includegraphics[width=.45\textwidth]{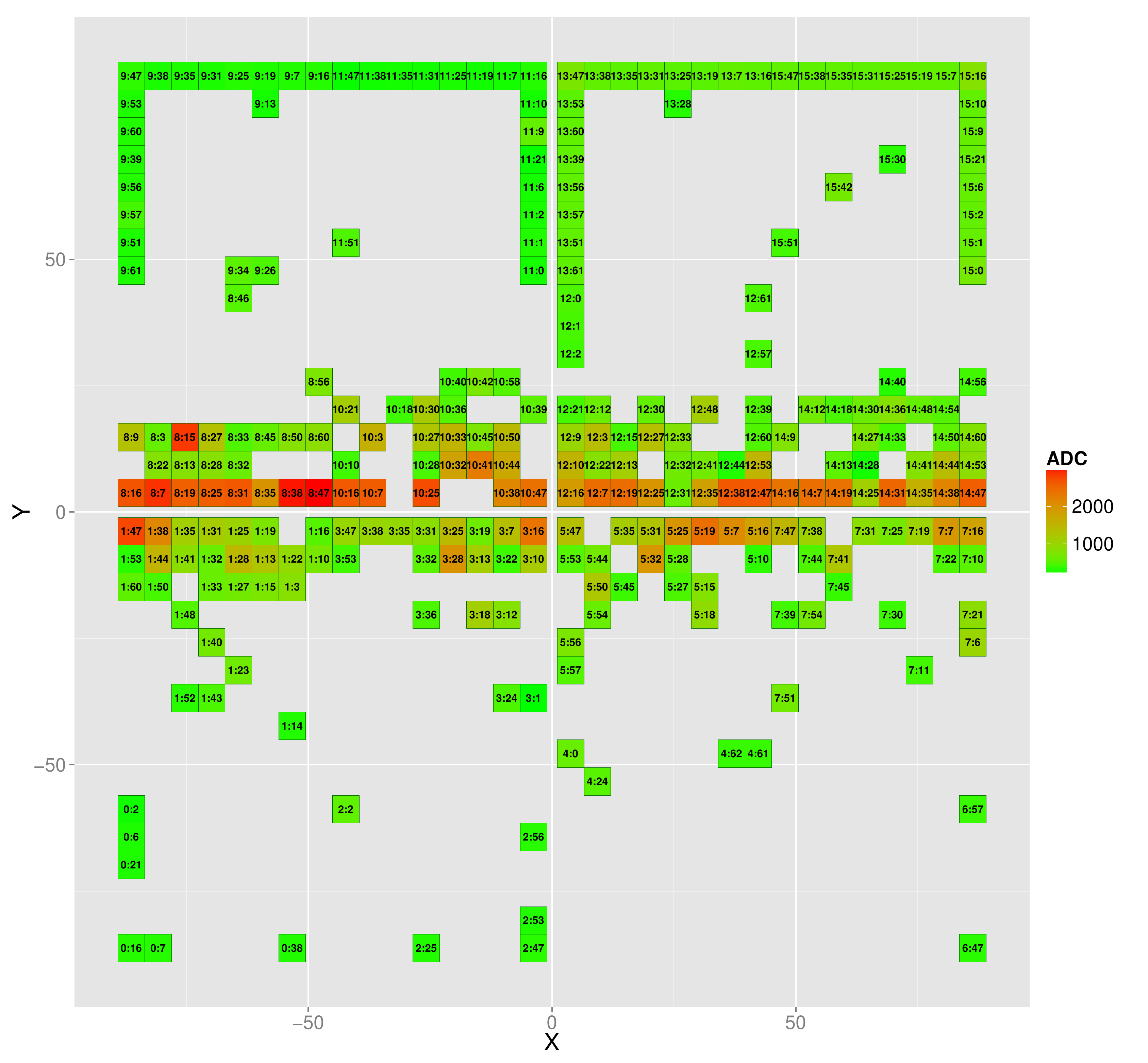}
	\vspace{-2.5em}
	\caption{$150GeV$ $\pi^{+}$ "square"-event example.}
	\label {squareevent}
	\vspace{-3.0em}
\end{wrapfigure}

Silicon sensor studies comprise:
\begin{itemize}
\item Silicon irradiation tests. Small sensors were irradiated by neutrons at Kobe tandem accelerator: $E_{n}<7.8MeV$, $d(3MeV)+Be(target)\rightarrow B+n(Q=4.36MeV)$. It was shown that increase of dark current is acceptable for 10 years of ILC operation at $\sqrt{s}=1TeV$ (Fig. \ref{darkcurrent}).
\item In the current design, a big energy deposition in the guard-ring at the sensor boundary can cause so-called "square"-events when almost all peripheral cells are fired. Small sensors with various guard-ring designs (1, 2, 4- or "no-GR") were produced and tested with laser injection in the corner of the sensor. $\sim 10 \%$ cross-talk between GR and peripheral cells is measured in the simplest 1-GR sensor. It is significantly suppressed in new no-GR and 2, 4-GR designs. For standard $9 \times 9 cm^{2}$ size sensors with improved GR design this effect was checked during November'15 tests at CERN. With the beam shooting in the center of the PCB, i.e. in the corners of 4 sensors, one can see very low rate of induced "square"-events ($<0.04\%$ for $150GeV$ $e^{+}$ showers after $8.4X_{0}$). Only with the beam shooting along GR side with ECAL turned by $90 \degree$, we got significant amount of "square"-events (Fig. \ref{squareevent}), proving that we are sensitive to the effect. 
\end{itemize}

\section{Future plans}

Currently SiW ECAL group is working on solutions, suitable for industrialization and mass production. Thanks to the power pulsing operation, a passive cooling is sufficient. For the ECAL this was checked with thermal studies. New version of VFE chip SKIROC2A is produced and is going to be tested. New ILD-like long detector from several ASUs is under development. There is an ongoing activity on alternative naked dye Chip On Board design, this technology allows to produce thinner detector, but it is rather fragile and difficult technologically.

This project has received funding from the European Union's Horizon 2020 Research and Innovation programme under Grant Agreement no. 654168.

\end{document}